\documentclass[preprint2]{aastex}

\newcommand{\Htwo}{$\rm{H_2}$~}
\newcommand{\Gnaught}{$G_0$}

\newcommand{\HIunits}{$10^{21}$ cm$^{-2}$}
\newcommand{\UVunits}{$10^{-15}$ ergs cm$^{-2}$ s$^{-1}$ \AA$^{-1}$}
\newcommand{\dtg}{$\delta/\delta_0$}

\slugcomment{To appear in ApJ}
\shorttitle{PDR-produced HI in M81}
\shortauthors{Heiner et al.}

\begin{document}

\title{The volume densities of GMCs in M81}

\author{Jonathan S. Heiner\altaffilmark{1,2}, Ronald J. Allen\altaffilmark{1}, Bjorn H.C. Emonts\altaffilmark{2,3}, Pieter C. van der Kruit\altaffilmark{2}
}
\altaffiltext{1}{Space Telescope Science Institute, Baltimore, MD 21218, USA}
\altaffiltext{2}{Kapteyn Astronomical Institute, University of Groningen, PO Box 800, 9700 AV Groningen, the Netherlands}
\altaffiltext{3}{Department of Astronomy, Columbia University, Mail Code 5246, 550 West 120th Street, New York, NY 10027, USA}
\email{heiner@stsci.edu}

\begin{abstract}
HI features near young star clusters in M81 are identified as the photodissociated surfaces of Giant Molecular Clouds (GMCs) from which the young stars have recently formed. The HI column densities of these features show a weak trend, from undetectable values inside $R = 3.7$\ kpc and increasing rapidly to values around $3 \times 10^{21}\ \mbox{cm}^{-2}$ near $R \approx 7.5$\ kpc. This trend is similar to that of the radially-averaged HI distribution in this galaxy, and implies a constant area covering factor of $\approx 0.21$ for GMCs throughout M81. The incident UV fluxes \Gnaught\ of our sample of candidate PDRs decrease radially.

A simple equilibrium model of the photodissociation-reformation process connects the observed values of the incident UV flux, the HI column density, and the relative dust content, permitting an independent estimate to be made of the total gas density in the GMC. Within the GMC this gas will be predominantly molecular hydrogen. Volume densities of $1 < n < 200\ \mbox{cm}^{-3}$ are derived, with a geometric mean of $17\ \mbox{cm}^{-3}$. These values are similar to the densities of GMCs in the Galaxy, but somewhat lower than those found earlier for M101 with similar methods. Low values of molecular density in the GMCs of M81 will result in low levels of collisional excitation of the CO(1-0) transition, and are consistent with the very low surface brightness of CO(1-0) emission observed in the disk of M81.
\end{abstract}


\section{Introduction}
\label{sec:intro}

Giant molecular clouds (GMCs) in the interstellar medium (ISM) are generally accepted as the birthplaces of new stars. The most massive of these new stars produce copious amounts of far ultraviolet (FUV) radiation which will, in turn, photodissociate the molecular gas of the parent GMCs, producing ``blankets'' of HI (and other atomic species) on their surfaces.  \citet{all1986} were the first to present evidence that major features in the HI distribution of the nearby spiral galaxy M83 (NGC 5236), namely, the inner HI spiral arms observed with the Very Large Array (VLA), were the result of photodissociation of \Htwo on galactic scales. \citet{all1997} confirmed that HI features existed in M81 (NGC 3031) on scales of $\approx 150$ pc which were qualitatively consistent with the expected morphology of large, low density photodissociation regions (PDRs), and explicitly related those HI features to nearby bright sources of Far-UV (FUV) radiation found on images of the galaxy from the Ultraviolet Imaging Telescope (UIT).

\citet{smi2000} used a simple, but quantitative, model for the equilibrium physics of photodissociation regions and applied it to VLA-HI and UIT-FUV observations of M101 (NGC 5457). Their work showed that this approach provides an entirely new method for determining the volume density of molecular gas in star-forming GMCs of galaxies, a method which is no longer dependent on the (often poorly known) excitation conditions for line emission by specific molecular tracers.

In this paper we return to another spiral galaxy, M81, and carry out a quantitative analysis of the GMCs in this galaxy using the methods described by \citet{smi2000}. This new analysis has been made possible by new data on M81 which was not available to \citet{all1997}; new, higher-resolution, high-sensitivity VLA-HI observations, and sensitive new FUV imagery from the GALEX satellite.  We have also sought an independent verification that the very basis of our approach is valid, namely, that the HI in the immediate vicinity of FUV concentrations is indeed produced in PDRs. To this end we have examined the Spitzer/IRAC data on M81 for evidence of mid-IR emission by polycyclic aromatic hydrocarbons (PAHs) which are also thought to be tracers of PDRs.

M81 has been probed extensively for the molecular tracer CO. No global CO map of M81 has been published to this date, and the detected CO emission is found to be very weak and spotty (see e.g. \citet{kna2006} and references therein). Since PDRs also produce CO emission \citep[e.g.][]{all2004}, comparing the CO results to the results in this paper is therefore of considerable interest.

The outline of this paper is as follows: Section \ref{sec:data} contains a description of the data we used, followed by a brief theoretical description of PDRs in Section \ref{sec:theory} and the application of the method in Section \ref{sec:method}. The results are presented in Section \ref{sec:results}. The results are discussed and the conclusions are briefly summarized in Section \ref{sec:discussionconclusions}.

\section{Data}
\label{sec:data}

\begin{figure}[tb]
  \plotone{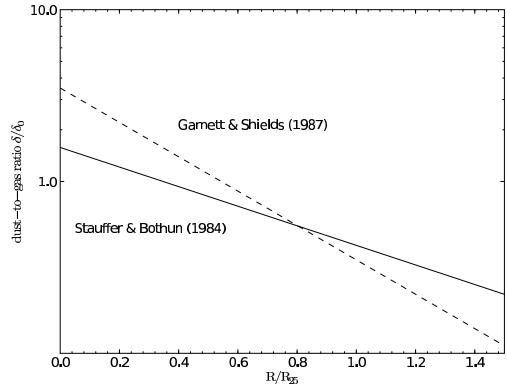}
  \caption{Dust-to-gas ratios as a function of $R$ in M81, as derived from \citet{gar1987} (dashed line) and \citet{sta1984} (solid line). $\delta_0$ is the value for the solar neighborhood in the Galaxy. We use the Stauffer \& Bothun data in this paper.}
  \label{fig:dust}
\end{figure}

\begin{figure}[tb]
  \epsscale{.90}
  \plotone{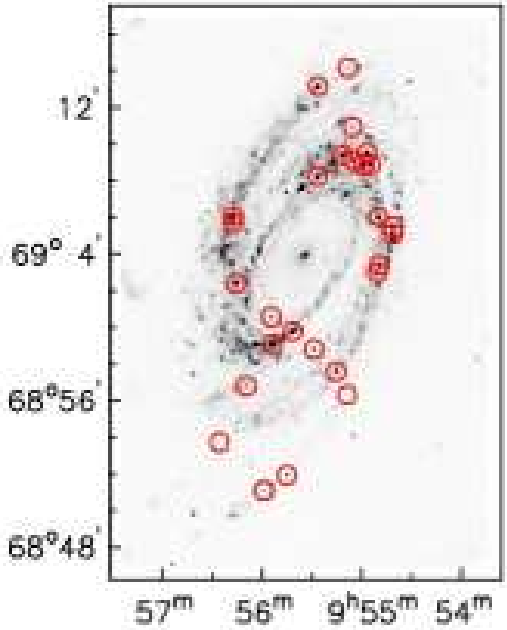}
  \caption{Locations of our 27 sources overlayed on the GALEX FUV image of M81.}
  \label{fig:locplot}
\end{figure}

\begin{figure*}[p]
  \includegraphics{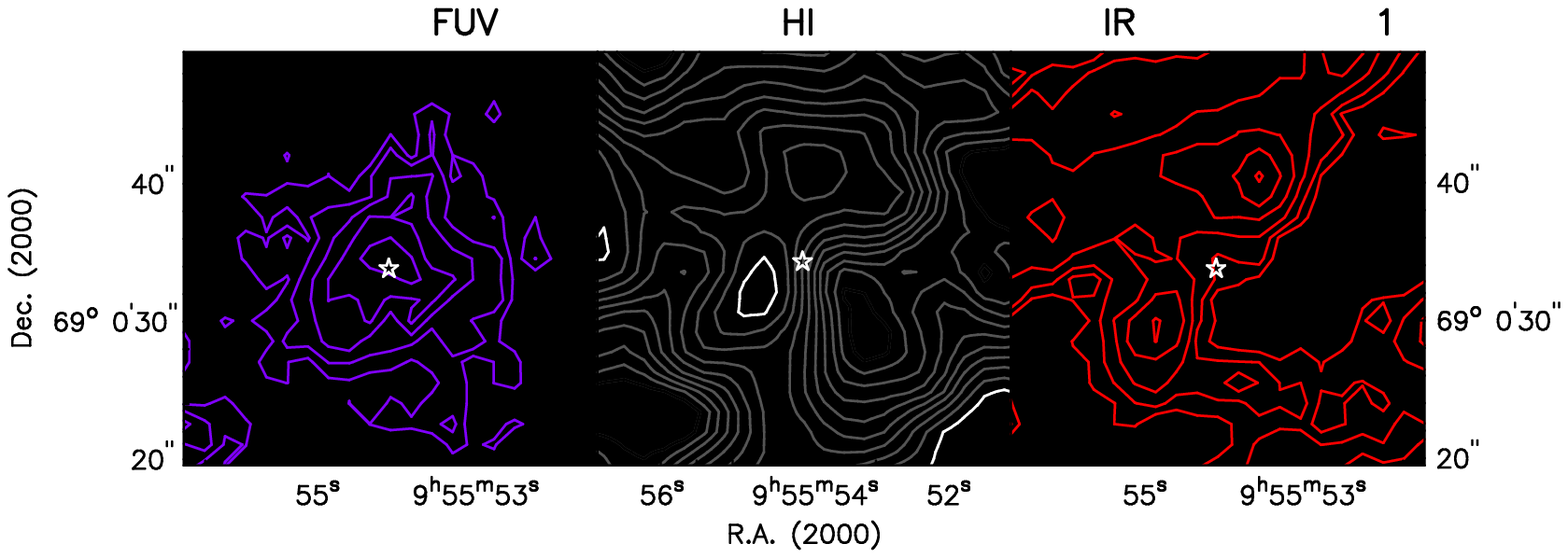}
  \includegraphics{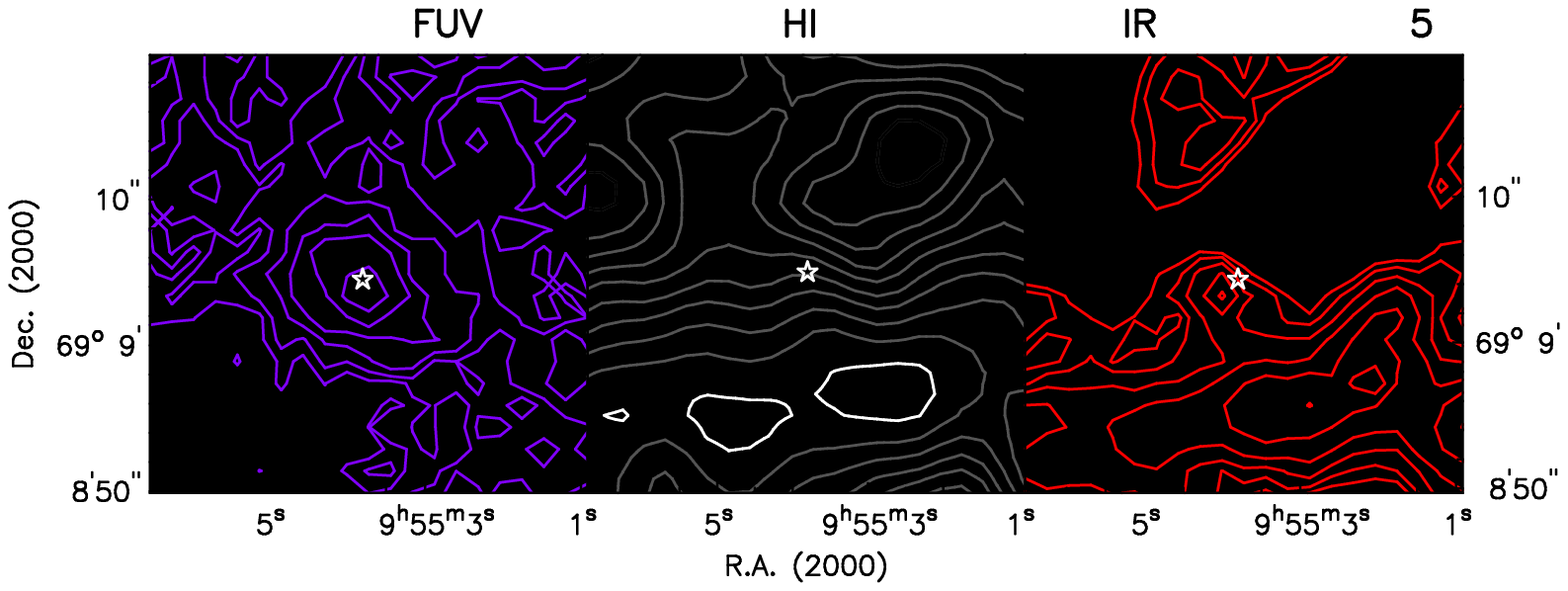}
  \caption{Detail plots of four sources from Table \ref{tab:FUVfluxes}. Their galactocentric radius is 4.1, 6.1, 7.8 and 9.6 kpc respectively. All sources are included in the electronic version. The HI is shown as the underlying grayscale in every panel. The central panel also shows HI contours. The lowest and highest contours are colored black and white respectively. The UV source (star) is marked in every panel. Left panel: UV contours (purple) start at $6 \times 10^{-18}$ ergs cm$^{-2}$ s$^{-1}$ \AA$^{-1}$, with each consecutive contour 1.48 times the previous one. The HI grayscale is as in the middle panel, but at the original pixel size of 1.5\arcsec. Middle panel: HI grayscale following the contours, of which the starting contour level and increment (previous level times increment) are given in Table \ref{tab:FUVfluxes}. Right panel: Non-stellar (PAH, red) contours at relative levels, starting at 1.25\%, with each consecutive contour 1.58 times the previous one. The HI grayscale is as in the middle panel, but at the original pixel size. Note the HI surrounding the UV sources in 'blankets', which becomes clearer after zooming out \citep{all1997}.}
  \label{fig:panels}
\end{figure*}

\begin{figure*}[p]
  \includegraphics{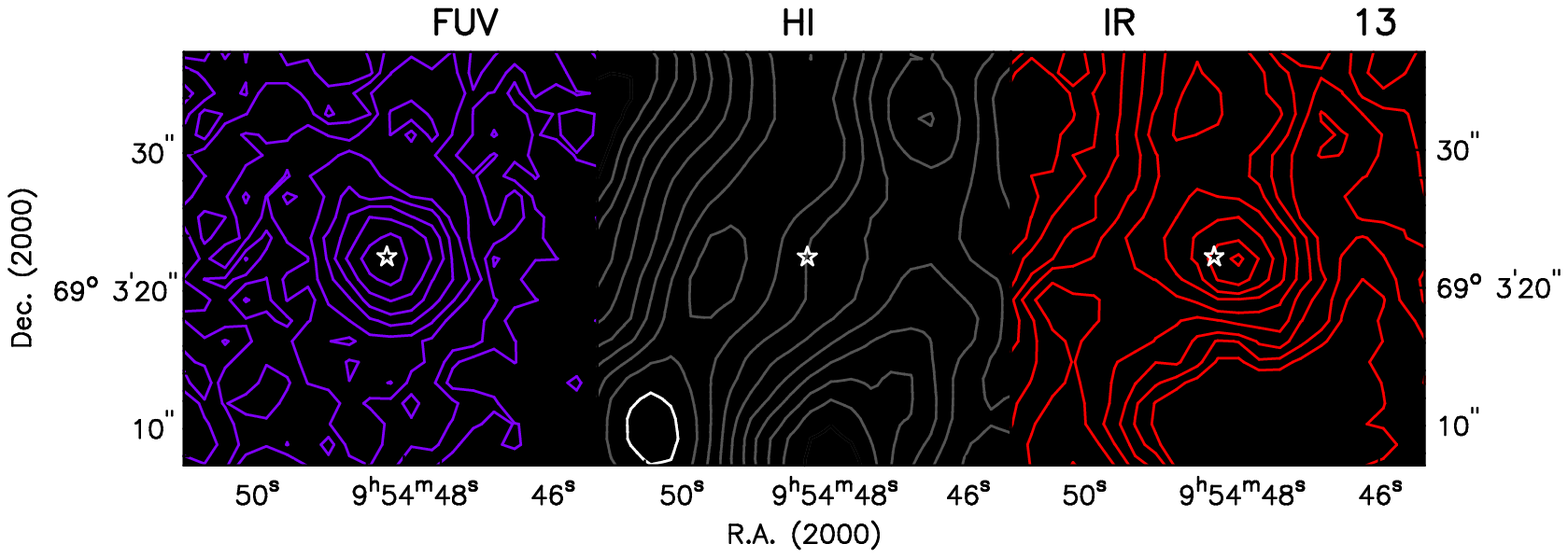}
  \includegraphics{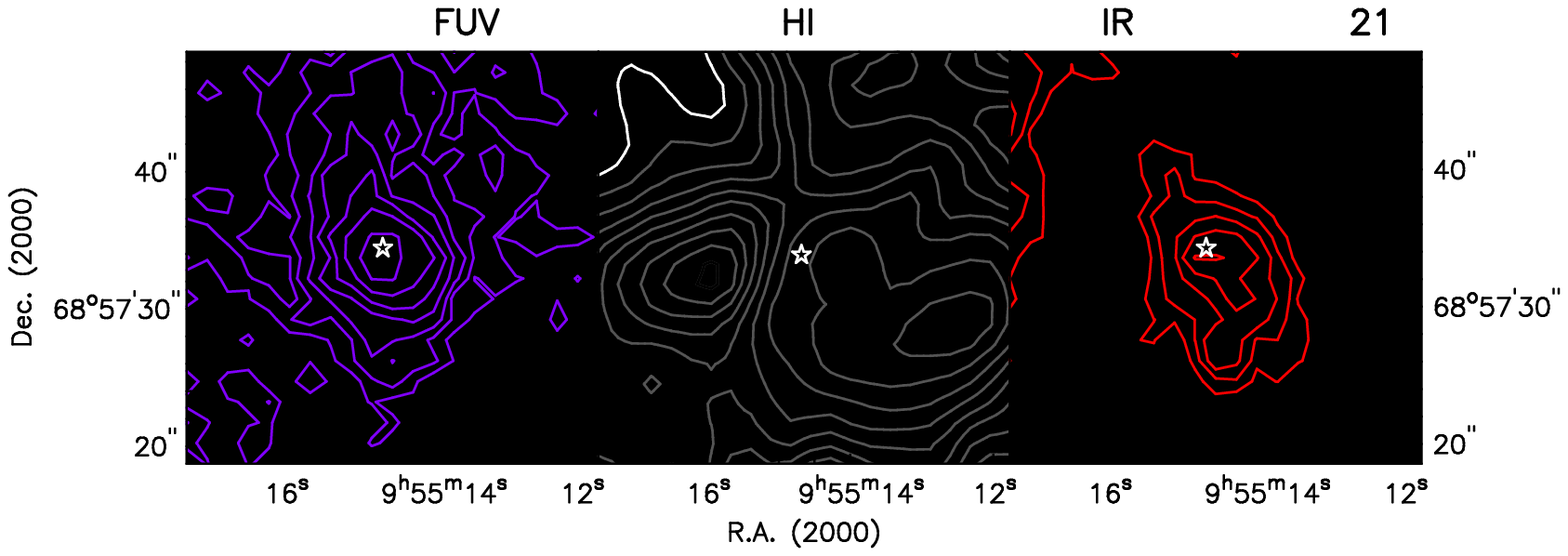}
  Figure \ref{fig:panels} - continued.
\end{figure*}

\begin{figure*}[tb!]
  \centering
  \includegraphics[width=0.8\textwidth]{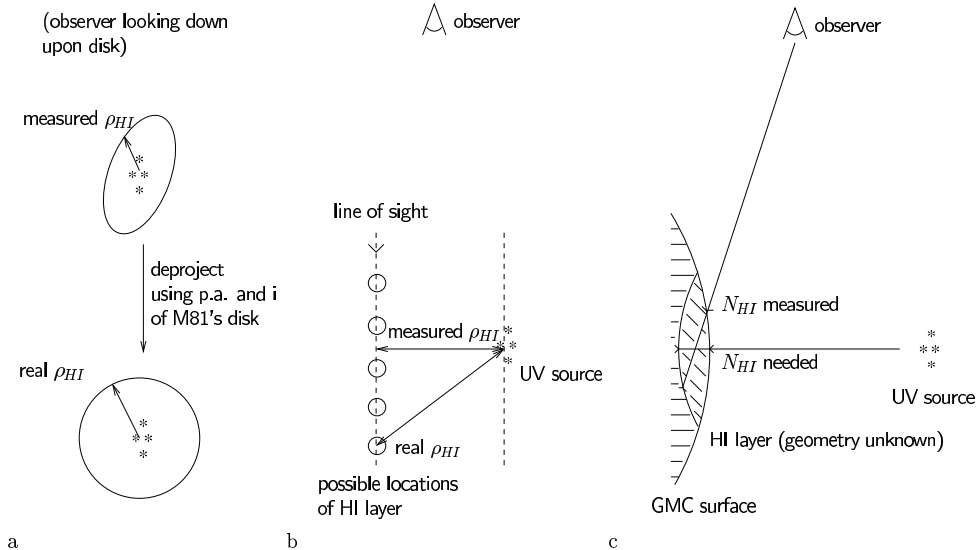}
  \caption{HI projection issues. Panel a: M81's position angle and inclination are used to deproject $\rho_{HI}$. Panel b: the exact vertical position of the HI patch along the line of sight is unknown. Panel c: the measured HI column is assumed to be a sufficient measure of the HI column needed for the PDR model, although the geometry of the HI layer is unknown.}
  \label{fig:HImethods}
\end{figure*}

\begin{figure}[t]
  \plotone{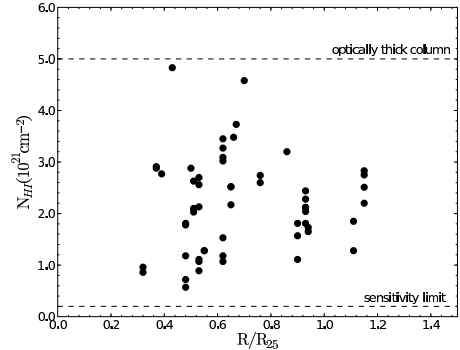}
  \caption{Individual (not ring-averaged) HI column densities from the THINGS M81 data. An approximate sensitivity limit is indicated. HI columns may become optically thick at $\approx 5\ \times$ \HIunits\ \citep{all2004}.}
  \label{fig:Rnorm_HI}
\end{figure}

\begin{figure}[t]
  \plotone{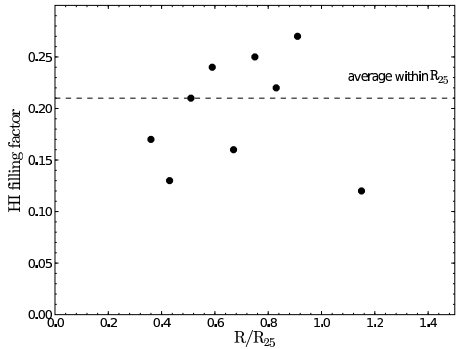}
  \caption{M81 filling factor as a function of galactocentric radius. The maximum HI column per kpc bin, converted to $M_\odot$ pc$^{-2}$, is divided by the ring-averaged values of M81's radial profile. The average filling factor within $R_{25}$ is 0.21. There is no data for $R/R_{25} < 0.3$.}
  \label{fig:Rnorm_FF}
\end{figure}

\begin{figure}[t]
  \plotone{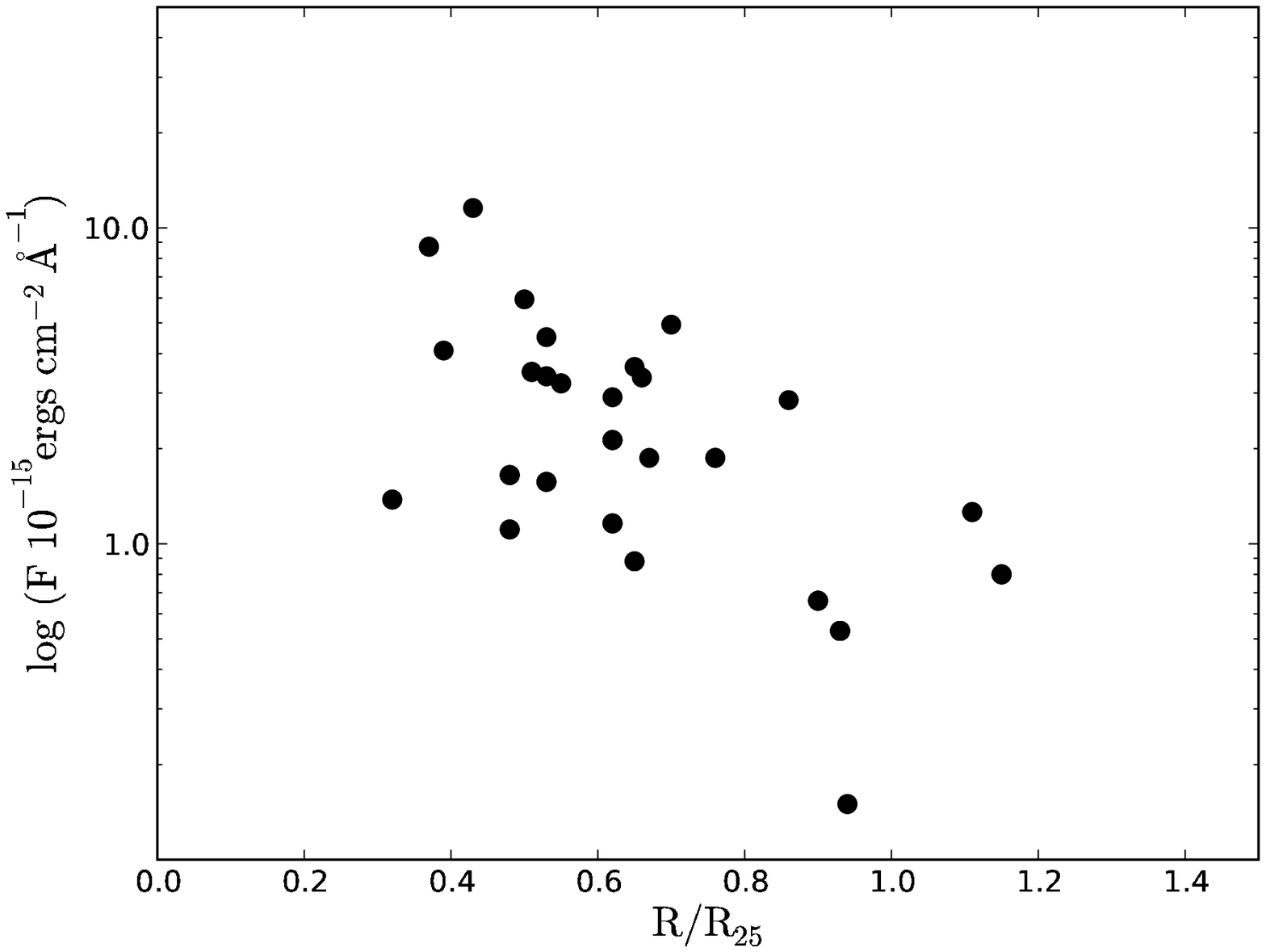}
  \caption{FUV source fluxes in \UVunits.}
  \label{fig:Rnorm_F}
\end{figure}

\begin{figure}[t]
  \plotone{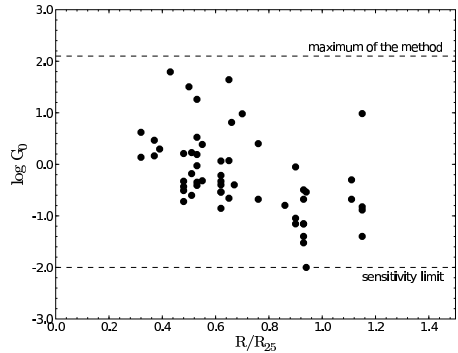}
  \caption{Incident flux \Gnaught\ on the HI patches chosed for study near the FUV sources.} 
  \label{fig:Rnorm_G0}
\end{figure}

\begin{figure}[t]
  \plotone{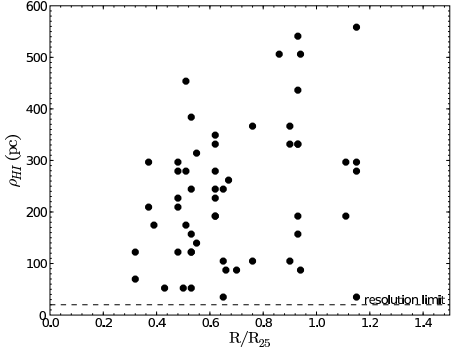}
  \caption{The distances $\rho_{HI}$ from the UV sources to any surrounding patches of HI.}
  \label{fig:Rnorm_rho}
\end{figure}

\begin{figure}[t]
  \plotone{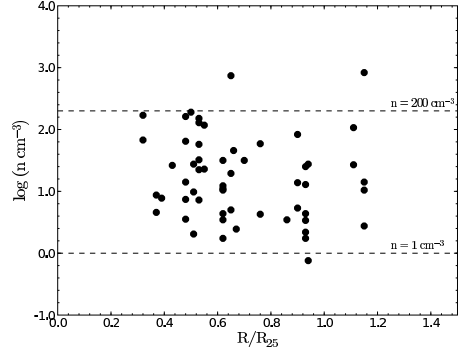}
  \caption{Derived total hydrogen volume densities of GMCs in M81. ($n = n_{HI} + 2n_{H_2} (\rm{cm^{-3}})$)}
  \label{fig:Rnorm_n}
\end{figure}

\begin{figure}[t]
  \plotone{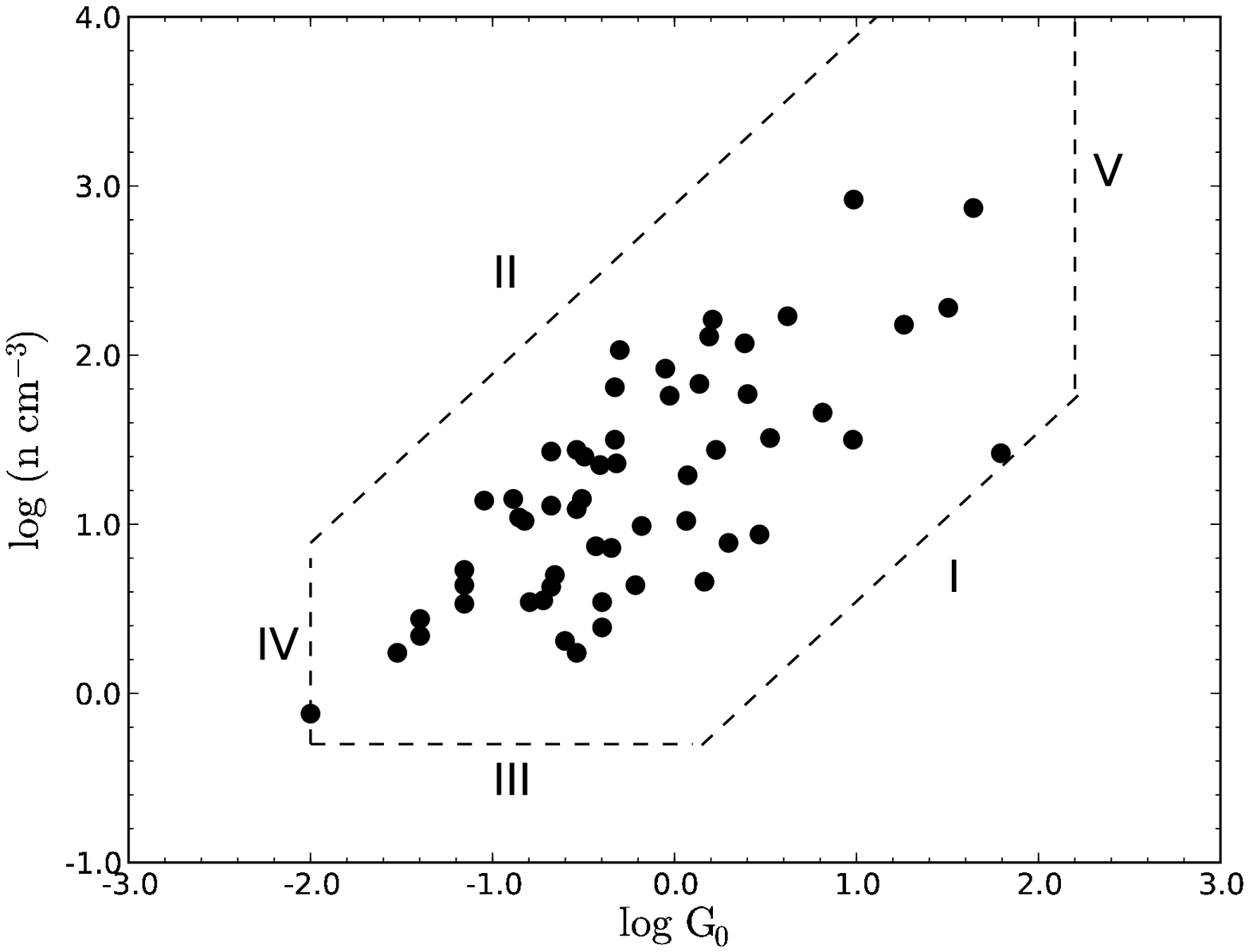}
  \caption{Various observational limits are shown in this plot of \Gnaught\ against $n$. The numerals are keyed to observational selection effects discussed in the text.}
  \label{fig:selection}
\end{figure}

The distance to M81 is taken to be 3.6 Mpc, after \citet{fre1994}, with an inclination of i = 58$\degr$ and position angle P.A. = 157$\degr$ \citep{vau1991}. At this distance, 1\arcsec\ = 17.5 pc in the plane of the sky. The galactocentric radii in all the plots have been normalized to $R_{25}$, which is 12.66 kpc for M81 \citep{vil1992}. For the model we used (Section \ref{sec:theory}), we need HI data, FUV data and the dust-to-gas ratio scaled to the local solar neighborhood value (\dtg).

A new, high quality integrated HI column density map of M81 made with the VLA has been made available by the THINGS team \citep{wal2005}. It has an angular resolution of 7.58\arcsec\ x 7.45\arcsec, equivalent to about a 130 pc linear resolution. The pixel spacing is 1.5\arcsec. An azimuthally-averaged radial profile of this data has been made available by Frank Bigiel (THINGS, private communication).

The ultraviolet data for M81 has been obtained from the publicly accessible GALEX data archive.  The angular resolution is 4\arcsec\ at a pixel spacing of 1.5\arcsec.  The astrometry has been checked against a number of stars in the Guide Star Catalogue v2.3. The Near-UV image was used for this purpose, since not all of these stars are detected in the Far-UV image. The NUV and FUV images were verified to have the same astrometry.
We use the FUV image at $\bar{\lambda} \approx 150 \mbox{\ nm}$ as a ``proxy'' measure of the photons with $\lambda \lesssim 100 \mbox{\ nm}$ that can directly dissociate molecular hydrogen, as was done in e.g. \citet{smi2000}.

The dust-to-gas ratio (\dtg) is derived from the metallicity measure $12 + \log(O/H)$. In the case of M81, the slope of log \dtg\ is taken as equal to the slope of $12 + \log(O/H)$ \citep{iss1990}. In the solar neighborhood, $12 + \log(O/H) = 8.8$, according to \citet{mih1981}. At this value, \dtg\ equals 1 by definition. This is sufficient to derive the expression we need for \dtg.

The $12 + \log(O/H)$ data for M81 has been adopted from \citet{sta1984}, but we also considered the results of \citet{gar1987}. The gradient they found is likely to be on the high side (Garnett, private communication). Figure \ref{fig:dust} shows both \dtg\ fits and illustrates the possible spread in the gradient of \dtg, similar to the spread in $12 + \log(O/H)$ that is implied from Figure 4 of \citet{gar1987}.
The fit to the \citet{sta1984} data we used yields 
\begin{equation}
\log(\delta/\delta_0) = -0.045 R + 0.198,
\label{eqn:dd0}
\end{equation}
where $R$ is the relevant galactocentric radius in kpc for each region under consideration. 

In order to identify PDRs from their PAH emission, we used M81 Spitzer data from the SINGS public release \citep{ken2003}. In particular, we used their 3.6 $\mu$m and 8.0 $\mu$m data from the IRAC camera. To verify that this data lined up correctly with our other data, we checked the astrometry against the same GSC stars as we used for the UV data and found the astrometry to be accurate to within 1\arcsec.

\section{Theory}
\label{sec:theory}

We use the data described in the previous section to derive the desired physical quantity $n$, the total density of gas in the GMC; $n = n_{HI} + 2n_{H_2} (\rm{cm^{-3}})$. The atomic hydrogen column density, FUV flux and dust-to-gas ratio are the required parameters to calculate the total hydrogen volume density.

Atomic hydrogen is slowly but continuously converted to molecular hydrogen in the interstellar medium, a process that is catalyzed by dust grains. Molecular hydrogen is rapidly photodissociated back into atomic hydrogen by UV photons.  Simple models for equilibrium concentrations of these components have been discussed by several authors, e.g. \citet{ste1988}, \citet{wol1990} and \citet{hol1991}.

\citet{all2004c} has reviewed the physics and compared the time scale for equilibrium with other time scales in galactic disks. His Equation 6 as derived from \citet{ste1988} explicitly includes the dependence on the dust-to-gas ratio, as well as improved values of the coefficients as determined by \citet{all2004} in their Equation A2. The equation we shall use from \citet{all2004c} is:

\begin{eqnarray}
\lefteqn{ N_{HI} = }\nonumber \\
& \frac{7.8 \times 10^{20}}{\delta/\delta_0} \ln\left[1+\frac{106G_0}{n}\left(\frac{\delta}{\delta_0}\right)^{-1/2}\right] \rm{cm^{-2}},
\label{eqn:NHI}
\end{eqnarray}
where \dtg\ is the dust-to-gas ratio scaled to the solar neighborhood value. $N_{HI}$ (in cm$^{-2}$) is the HI column produced by photodissociation.  $n = n_{HI} + 2n_{H_2} (\rm{cm^{-3}})$ is the total hydrogen volume density; this gas is mostly atomic on the surface of the GMC and mostly molecular deep inside the cloud.  \Gnaught\ is a dimensionless parameter measuring the strength of the incident UV field. It is scaled to twice the equivalent Habing flux (at 1500 \AA), which is appropriate for an FUV-opaque PDR illuminated over $2\pi$ sr \citep{hol1999,all2004}. The latter authors also discuss how the flux and its scaling at 1500 \AA\ relate to the 1000 \AA\ flux, which is in the primary range of energies of dissociation photons. \Gnaught\ is equivalent to 0.85 times the field strength defined in \citet{dra1978} and used by \citet{ste1988}. It does not include the ambient UV background radiation, so it is calculated from a background subtracted UV flux. We evaluate \Gnaught\ as follows:

\begin{equation}
G_0 = F_{FUV}\left(D_{gal} / \rho_{HI}\right)^2/F_0,
\label{eqn:G0}
\end{equation}
where $D_{gal}$ is the distance of M81 and $\rho_{HI}$ is the distance between the UV source and the location of the atomic hydrogen column. $F_0 = 2 \times F_H = 3.11 \times 10^{-6}\ \mbox{ergs cm}^{-2}\ \mbox{s}^{-1}\ \mbox{\AA}^{-1}$ at 1500 \AA\ \citep{all2004,dis1988}.

Equation \ref{eqn:NHI} can be inverted in order to obtain the total hydrogen volume density $n$, using $N_{HI}$, \Gnaught\ and $\delta/\delta_0$. The resulting equation is:

\begin{eqnarray}
n &=& \frac{106G_0}{(\delta/\delta_0)^{1/2}} \times \nonumber \\
&& \left[\exp{\lbrace N_{HI}(\delta/\delta_0)/7.8\times10^{20}\rbrace}-1\right]^{-1}.
\label{eqn:n}
\end{eqnarray}
Equation \ref{eqn:n} is valid for PDRs with $n < 10^4\ \mbox{cm}^{-3}$. 

The morphology of the regions identified by \citet{all1997} is that of HI patches in the form of shells (or parts of shells) partially surrounding young star clusters. Although this morphology appears ubiquitous in M81 and other galaxies, independent confirmation of these HI features as PDRs would be very desirable. Such a confirmation could be provided by the presence of PAH emission. Although no quantitative relation between PAH, UV and HI emission at this scale is currently available, the detection of PAHs is thought to be predominantly linked to UV emission \citep{uch1998,lid2002}.

\section{Method}
\label{sec:method}

\begin{deluxetable}{rccccccc}
\tabletypesize{\scriptsize}
\tablecaption{Locations and FUV fluxes of candidate PDRs.}
\tablehead{ 
  \colhead{Source} &
  \colhead{R.A.(2000)} &
  \colhead{DEC(2000)} &
  \colhead{Radius} &
  \colhead{F$_{FUV}$} &
  \colhead{Aperture} &
  \colhead{Cont. start} &
  \colhead{Increment}\\ 
  \colhead{no.} &
  \colhead{(h m s)} &
  \colhead{(d m s)} &
  \colhead{(kpc)} &
  \colhead{(\UVunits)} &
  \colhead{(\arcsec)} &
  \colhead{($10^{20}$ cm$^{-2}$)} &
  \colhead{($10^{20}$ cm$^{-2}$)}
}  
\startdata   
  1 \ \ &9 55 54.016  & 69 00 33.79 & \phn4.1 & \phn1.38 & 17 & \phn5.0 & 1.080 \\
  2 \ \ &9 55 41.057  & 68 59 45.82 & \phn4.7 & \phn8.72 & 20 & \phn7.5 & 1.170 \\
  3 \ \ &9 55 25.268  & 69 08 15.24 & \phn4.9 & \phn4.09 & 10 & \phn5.0 & 1.210 \\
  4 \ \ &9 55 53.182  & 68 59 04.74 & \phn5.4 &11.56 & 24 & \phn9.0 & 1.190 \\
  5 \ \ &9 55 03.784  & 69 09 04.41 & \phn6.1 & \phn1.11 & 10 & \phn6.5 & 1.120 \\
  6 \ \ &9 55 10.104  & 69 09 21.54 & \phn6.1 & \phn1.65 & 10 & \phn5.0 & 1.095 \\
  7 \ \ &9 54 57.442  & 69 08 49.19 & \phn6.3 & \phn5.94 & 15 & \phn6.5 & 1.170 \\
  8 \ \ &9 56 15.363  & 69 02 23.58 & \phn6.4 & \phn3.50 & 16 & \phn4.5 & 1.220 \\
  9 \ \ &9 54 48.655  & 69 06 00.02 & \phn6.7 & \phn3.39 & 13 & \phn9.0 & 1.130 \\
 10 \ \ &9 55 28.266  & 68 58 47.66 & \phn6.7 & \phn1.57 & 13 & \phn4.0 & 1.120 \\
 11 \ \ &9 54 53.138  & 69 08 51.02 & \phn6.8 & \phn4.51 & 15 & \phn5.0 & 1.180 \\
 12 \ \ &9 54 55.272  & 69 09 28.97 & \phn7.0 & \phn3.22 & 13 & \phn4.5 & 1.130 \\
 13 \ \ &9 54 48.210  & 69 03 22.35 & \phn7.8 & \phn2.91 & 16 &11.0 & 1.130 \\
 14 \ \ &9 55 03.708  & 69 10 54.55 & \phn7.8 & \phn1.16 & 12 & \phn7.7 & 1.080 \\
 15 \ \ &9 54 49.476  & 69 02 56.82 & \phn7.9 & \phn2.13 & 17 & \phn9.5 & 1.140 \\
 16 \ \ &9 56 09.346  & 68 56 48.16 & \phn8.2 & \phn0.88 & 13 & \phn6.0 & 1.175 \\
 17 \ \ &9 54 39.443  & 69 05 26.60 & \phn8.3 & \phn3.63 & 15 & \phn7.5 & 1.150 \\
 18 \ \ &9 54 39.949  & 69 05 03.44 & \phn8.3 & \phn3.36 & 12 &13.0 & 1.115 \\
 19 \ \ &9 56 17.179  & 69 05 42.48 & \phn8.5 & \phn1.87 & 13 &20.0 & 1.080 \\
 20 \ \ &9 56 17.148  & 69 06 04.85 & \phn8.8 & \phn4.94 & 17 &14.0 & 1.135 \\
 21 \ \ &9 55 14.667  & 68 57 34.35 & \phn9.6 & \phn1.87 & 13 &15.0 & 1.075 \\
 22 \ \ &9 55 25.114  & 69 13 09.70 & 10.9 & \phn2.85 & 13 & \phn7.5 & 1.150 \\
 23 \ \ &9 55 06.623  & 69 14 08.97 & 11.4 & \phn0.66 & 15 & \phn6.0 & 1.130 \\
 24 \ \ &9 56 25.503  & 68 53 43.61 & 11.8 & \phn0.53 & 15 & \phn9.5 & 1.095 \\
 25 \ \ &9 55 08.312  & 68 56 17.30 & 11.9 & \phn0.15 & \phn6  & \phn5.5 & 1.135 \\
 26 \ \ &9 55 44.892  & 68 51 54.80 & 14.1 & \phn1.26 & 12 & \phn4.0 & 1.170 \\
 27 \ \ &9 55 58.783  & 68 51 04.79 & 14.5 & \phn0.80 & 12 & \phn5.0 & 1.210 \\
\enddata	
\label{tab:FUVfluxes}	 	          
\end{deluxetable}

In this Section, we explain how the total hydrogen volume density in every candidate PDR is determined by investigating the HI structures surrounding the central FUV source. We hypothesize that this source (a cluster of young stars) creates a photon dominated region, in which molecular hydrogen is converted into atomic hydrogen. While the exact morphology of these regions remains unclear, the distribution of HI suggests that large scale photodissociation is taking place. The presence of PAH emission in these regions reinforces this picture.

We selected 27 sources based on their well-defined compact FUV emission, from faint to bright. The sources were chosen to have a range of galactocentric radii, although we suspect that all candidate PDRs should have similar total hydrogen densities regardless of their location \citep{smi2000}. It is significant that we found no sources of this type within $R/R_{25} \approx 0.3$ and brighter than our faintest source of $0.15 \times$ \UVunits. There is one source at a smaller radius that has a similar FUV flux as our faintest source, but it is extended and has a lower contrast than our faintest source. The locations of these sources are shown in Figure \ref{fig:locplot} and given in Table \ref{tab:FUVfluxes}.

Four example sources are shown in Figure \ref{fig:panels}. These sources vary in FUV flux and galactocentric distance. Each source shows the general structure of a UV source surrounded or bordered by HI with PAH emission nearby. At larger scales this is clearly visible, see \citet{all1997}, or \citet{smi2000} for similar plots of M101.

The FUV source flux is determined first. We correct for galactic foreground extinction, but not for internal extinction. We decided against applying an internal extinction correction, despite the availability of internal extinction data \citep{hil1995}. It is assumed that the height of M81's gas disk is of the same order as the size of the PDRs to which our method is most sensitive. \citet{vis1980} use a gas disk half height of up to 450 pc, which is comparable to the size of our candidate PDRs. In that case the extinction from the HI patch to the UV source is the same as the extinction of the UV source in M81's disk towards the line of sight of the observer. Under these circumstances, the corresponding extinction corrections to the UV flux cancel each other out. Using the internal extinction correction is therefore unnecessary.

The radius at which the UV emission blends into the background radiation level determines the size of the UV source on the image and the place to measure the background. The total FUV flux within this radius is calculated and the background flux is subtracted. Typically, the net flux is 50\% of the total FUV flux. For comparison, we calculate the contrast of the source above the background level, which we define as the incident UV flux on the HI patches $G$ (the unscaled \Gnaught) divided by the incident background UV flux $G_{bg}$. The latter is defined as the UV background surface brightness at the edge of the aperture of the UV source, integrated over half the sky (a flux on a semi-infinite slab).  Only the net flux of the source is used to calculate \Gnaught, since the model does not include calculating HI produced by the ambient UV background. The measured flux is in ergs cm$^{-2}$\ s$^{-1}$\ \AA$^{-1}$. 

The distance $\rho_{HI}$ from the UV source position to any surrounding patches (concentrations) of HI is measured next, along with the column density of each such patch. A patch of HI is identified as a local maximum in column density. See Figure \ref{fig:panels}, middle panels. This procedure differs slightly from that of \citet{smi2000}, who took the first local maximum of the ring averaged HI column densities around UV sources as a probe of the true HI column density. In fact, any local bright spot in atomic hydrogen is an equally valid probe of a potential PDR. They are all likely to be direct views onto the surfaces of molecular clouds surrounding the UV source. In that case, the HI local maxima are a measure of the true local column density and each measurement represents (part of) an HI cloud. For this reason we decided to use more than one measurement of the HI surrounding the UV source. We subtract a general background level of HI emission that we assume has either not been created by photodissociation, or that has been created by the general UV background radiation field acting on all GMCs in the area.

In addition to identifying individual HI concentrations, the maximum measured column densities per 1 kpc galactocentric bin are recorded and compared to the ring-averaged values from M81's radial profile. Dividing the maximum column densities per bin of individual HI clouds, converted to $M_\odot$ pc$^{-2}$, by the average surface density per 1 kpc bin from M81's radial profile yields a filling factor. In the framework of the derived PDR model, this filling factor is an estimate of the extent to which GMCs fill M81's disk.

In order to determine $\rho_{HI}$, we corrected for M81's inclination and position angle assuming M81's disk to be intrinsically circular (see Figure \ref{fig:HImethods}a). From this, the incident flux \Gnaught\ on each surrounding HI patch is computed using Equation \ref{eqn:G0}. The relevant dust-to-gas ratio at the galactocentric radius of each UV source is taken from Figure \ref{fig:dust}. The extinction between M81 and the observer is taken to be 1.37 mag at 1500 \AA, after \citet{all1997} and \citet{sea1979}. The measured FUV flux is increased accordingly when calculating \Gnaught.  The incident flux \Gnaught, the HI column density and the dust-to-gas ratio \dtg\ finally yields a total hydrogen volume density for each patch according to Equation \ref{eqn:n}.  Note that this probe for molecular gas does not yield any information about the geometry of the gas clouds, but deep inside the GMC all available hydrogen is likely to be in the form of molecular hydrogen, so $n = 2n_{H_2}$.

A PAH map was obtained by subtracting a scaled and smoothed version of the 3.6 $\mu$m M81 image from the 8 $\mu$m M81 image, as described by \citet{wil2004}.  This procedure assumes that the 3.6 $\mu$m image contains purely stellar emission, while the 8 $\mu$m image contains a mix of stellar and PAH emission.  We determined the required scale factor empirically, by comparing four regions of M81 encompassing most of our candidate PDRs. Correlation maps of individual pixels were used to determine the slope of the appropriate correlated component.  Our four scale factors are consistent with the single one used by \citet{wil2004}, being within a range of 0.7 and 1. Seven sources were not included in these four regions; these sources have comparatively large galactocentric radii and any 3.6 $\mu$m contribution is negligible. In these cases we therefore used the raw 8 $\mu$m image.  The relevance of the detected PAH emission around the candidate PDR was tested by checking for the presence of PAH emission near the location of the UV source and determining the separation between PAH emission peaks and HI patches.

Summarizing: determining $N_{HI}$, $\rho_{HI}$, $F_{FUV}$ and \dtg\ yields $n$,  derived using the picture of photodissociation. Additionally, the presence of PAH emission reinforces the photodissocation assumption.

\section{Results}
\label{sec:results}

In this Section, we present total hydrogen volume densities of the 27 candidate PDRs. Table \ref{tab:FUVfluxes} lists the FUV fluxes of the central (cluster of) sources. Table \ref{tab:NHI} lists the corresponding HI column density measurements, the resulting incident fluxes and finally the total hydrogen volume densities. First, we'll discuss the HI measurements, followed by the FUV results. These results, combined with the appropriate dust-to-gas ratio from Equation \ref{eqn:dd0}, are then used to calculate $n$. At the end of this Section, we discuss the spatial correlation between HI and PAHs in M81.

\subsection{Atomic Hydrogen}
Each region can have one or more HI measurements, as was explained in the previous Section. The resulting HI column densities and their distance to the central FUV source are listed in Table \ref{tab:NHI}. 

Figure \ref{fig:Rnorm_HI} shows the HI column densities of all HI patches from Table \ref{tab:NHI} as a function of (normalized) galactocentric radius.  The general background column density level is of the order of 1 $\times\ 10^{20}$ cm$^{-2}$ and has been subtracted from all measured values, for reasons mentioned above (Section \ref{sec:method}).  Very little HI radio emission is observed within a radius of 3 kpc of the center of M81.  The HI column densities rise with increasing galactocentric radius from 0.3 to 0.6 $\times\ R_{25}$ (3.8 kpc to 7.5 kpc), although there is significant scatter.

While the general morphology of our candidate PDRs is the same (FUV emission surrounded by atomic hydrogen), the actual geometry of the GMCs can not be discerned at the available resolution. The HI structures are barely resolved, which makes it practically impossible to deal with projection effects properly. This introduces inaccuracies in the determination of $\rho_{HI}$ (Figure \ref{fig:Rnorm_rho}, discussed below) and $N_{HI}$. 

We can not include HI patches closer than about 20 pc (roughly the size of one pixel), or right on top of a UV source, even though they could have a significant separation in the vertical direction. The model can not take a $\rho_{HI}$ of 0. We had to reject only one HI patch because it was too close to the UV source. The upper limit of $\rho_{HI}$ (about 600 pc) is somewhat arbitrary and dependent on the maximum separation that could still conceivably have HI connected to a certain UV source. In combination with the lowest measured FUV flux, the resulting \Gnaught\ would be 0.01.

We corrected for M81's inclination and position angle (Figure \ref{fig:HImethods}a), assuming M81's disk to be perfectly circular in shape. This means we used ellipses to measure the deprojected separation between the FUV source and the associated HI patches. However, $\rho_{HI}$ is still sensitive to projection effects, as illustrated in Figure \ref{fig:HImethods}b. It shows how the measured separation could further be deprojected if we knew the position of all HI patches in the z-direction of the disk of M81. Therefore, the measured separation $\rho_{HI}$ is close to, but not exactly equal to the real separation. This will lead to underestimating the separation between the UV source and its associated HI patches. This in turn will mean overestimating \Gnaught\ and therefore overestimating $n$ as well. At the same time, due to our finite resolution, there is a lower limit to the separation we can observe. The lowest value of $\rho_{HI}$ we measured was 35 pc.

\begin{deluxetable}{rccrcrrc}
\tabletypesize{\scriptsize}
\tablecaption{HI measurements, incident fluxes and PAH correlation.}
\tablewidth{0pt}
\tablehead{
  \colhead{Source} & 
  \colhead{$\rho_{HI}$} &
  \colhead{$N_{HI}$} &
  \colhead{\Gnaught} & 
  \colhead{$G/G_{bg}$} &
  \colhead{$n$} &
  \colhead{Error} &
  \colhead{PAH-HI} \\
  \colhead{no.} &
  \colhead{(pc)} &
  \colhead{(\HIunits)} &
  \colhead{} &
  \colhead{} & 
  \colhead{(cm$^{-3}$)} &
  \colhead{(\%)} &
  \colhead{$<$ 6\arcsec apart}
}
\startdata
1a	 & \phn70 & 0.96  & 4.18  & \phn3.89 & 169 & 50  & n \\
 b	 & 122	  & 0.86  & 1.37  & \phn1.27 & 67  & 44  & n \\
2a	 & 209	  & 2.91  & 2.93  & \phn1.41 & 5   & 88  & y \\
b  	 & 297	  & 2.88  & 1.46  & \phn0.70 & 9   & 87  & y \\
3	 & 175	  & 2.77  & 1.98  & \phn0.47 & 8   & 85  & y \\
4	 & \phn52 & 4.83  & 62.12 & 15.20    & 27  & 140 & y \\
5a	 & 209	  & 1.78  & 0.37  & \phn0.17 & 7   & 66  & n \\
b  	 & 297	  & 1.81  & 0.19  & \phn0.08 & 4   & 66  & y \\
6a	 & 122	  & 0.72  & 1.62  & \phn0.29 & 163 & 48  & y \\
b   	 & 227	  & 0.57  & 0.47  & \phn0.08 & 64  & 44  & n \\
c   	 & 279	  & 1.18  & 0.31  & \phn0.05 & 14  & 54  & y \\
7	 & \phn52 & 2.88  & 31.93 & \phn8.70 & 192 & 96  & y \\
8a	 & 175	  & 2.03  & 1.69  & \phn0.63 & 27  & 72  & y \\
b  	 & 279	  & 2.10  & 0.66  & \phn0.25 & 10  & 73  & n \\
c  	 & 454	  & 2.63  & 0.25  & \phn0.09 & 2   & 85  & y \\
9a	 & \phn52 & 2.70  & 18.23 & \phn6.58 & 151 & 93  & y \\
b  	 & 122	  & 2.56  & 3.35  & \phn1.21 & 32  & 85  & y \\
10a	 & 122	  & 0.89  & 1.55  & \phn1.94 & 128 & 53  & y \\
b  	 & 157	  & 1.07  & 0.94  & \phn1.18 & 58  & 55  & y \\
c	 & 244	  & 1.11  & 0.39  & \phn0.49 & 23  & 55  & n \\
11 	 & 384	  & 2.13  & 0.45  & \phn0.15 & 7   & 75  & y \\
12a	 & 140	  & 1.28  & 2.43  & \phn0.99 & 117 & 60  & y \\
b  	 & 314	  & 1.28  & 0.48  & \phn0.19 & 23  & 59  & y \\
13	 & 192	  & 3.02  & 1.16  & \phn0.46 & 10  & 98  & n \\
14a	 & 192	  & 1.18  & 0.47  & \phn0.48 & 31  & 61  & n \\
b 	 & 244	  & 1.53  & 0.29  & \phn0.30 & 12  & 67  & n \\
c 	 & 349	  & 1.07  & 0.14  & \phn0.15 & 11  & 59  & y \\
15a  	 & 227	  & 3.27  & 0.61  & \phn0.25 & 4   & 104 & y \\
b    	 & 279	  & 3.09  & 0.40  & \phn0.17 & 3   & 99  & y \\
c     	 & 332	  & 3.45  & 0.29  & \phn0.12 & 2   & 107 & y \\
16a	 & 105	  & 2.52  & 1.18  & \phn0.92 & 19  & 89  & y \\
b  	 & 244	  & 2.17  & 0.22  & \phn0.17 & 5   & 81  & n \\
17	 & \phn35 & 2.52  & 43.88 & 16.23& 737 & 102 & y \\
18	 & \phn87 & 3.48  & 6.51  & \phn1.71 & 46  & 111 & y \\
19	 & 262    & 3.73  & 0.40  & \phn0.10 & 2   & 116 & y \\
20	 & \phn87 & 4.58  & 9.56  & \phn2.39 & 32  & 138 & y \\
21a	 & 105	  & 2.60  & 2.52  & \phn1.36 & 59  & 97  & y \\
b 	 & 367	  & 2.74  & 0.21  & \phn0.11 & 4   & 98  & n \\
22	 & 506	  & 3.20  & 0.16  & \phn0.05 & 3   & 113 & y \\
23a      & 105    & 1.57  & 0.89  & \phn1.44 & 83  & 87  & n \\
b        & 332    & 1.11  & 0.09  & \phn0.14 & 14  & 78  & n \\
c        & 367    & 1.81  & 0.07  & \phn0.12 & 5   & 90  & y \\\tableline
\tablebreak
24a      & 157    & 1.81  & 0.32  & \phn1.02 & 25  & 92  & n \\
b        & 192    & 2.12  & 0.21  & \phn0.68 & 13  & 97  & n \\
c        & 332    & 2.44  & 0.07  & \phn0.23 & 3   & 103 & n \\
d        & 332    & 2.12  & 0.07  & \phn0.23 & 4   & 97  & n \\
e        & 436    & 2.28  & 0.04  & \phn0.13 & 2   & 100 & n \\
f        & 541    & 2.04  & 0.03  & \phn0.09 & 2   & 96  & n \\
25a      & \phn87 & 1.65  & 0.29  & \phn0.50 & 28  & 92  & y \\
b        & 506    & 1.73  & 0.01  & \phn0.01 & 1   & 91  & n \\
26a	 & 192	  & 1.28  & 0.50  & \phn0.87 & 107 & 100 & n \\
b 	 & 297	  & 1.85  & 0.21  & \phn0.36 & 27  & 109 & n \\
27a      & \phn35 & 2.51  & 9.65  & 23.39& 828 & 134 & y \\
b        & 279    & 2.83  & 0.15  & \phn0.37 & 11  & 129 & n \\
c        & 297    & 2.20  & 0.13  & \phn0.32 & 14  & 118 & n \\
d        & 559    & 2.75  & 0.04  & \phn0.09 & 3   & 128 & n \\
\enddata
\label{tab:NHI}
\end{deluxetable}

Additionally, the HI column that is needed for Equation \ref{eqn:n} is defined as seen along the line of sight from the UV source to the HI layer (Figure \ref{fig:HImethods}c). The observer sees the HI column from a different angle. Since the geometry of the HI layer is unknown, we are assuming that the observed column is nevertheless a reasonable measure of the proper HI column.

The range of observable HI columns is indicated in Figure \ref{fig:Rnorm_HI}.  With a signal-to-noise ratio of $\frac{S}{N} \approx 4-5$, only regions with column density above $1\ \times$ \HIunits\ can be reliably distinguished as HI clouds. ($N_{HI}$ reaches this level in the region surrounding source no. 11.) This limits the maximum observable total hydrogen volume density.

At a column density of about $5\ \times$ \HIunits, the HI columns may become optically thick \citep{all2004}. In our sample, two sources have HI columns approaching this value. One of these sources has a relatively high UV flux. Neither source has an unusual total hydrogen volume density. 

Our method is biased towards low density, large scale HI regions. The HI from compact high density PDRs like the Orion Nebula in our Galaxy can not be currently detected at the distance of M81 because of the limited resolution of the HI map. The high density regions will be of small scale size generally and difficult to discern from the generally-lumpy HI background. Beam smoothing will also reduce the measured value of $N_{HI}$ in this case. Detectable quantities of HI can be produced through a combination of a high \Gnaught\ and high $n$ or a low \Gnaught\ and low $n$ (recall Equation \ref{eqn:NHI}). The highest \Gnaught\ we can detect is limited by the maximum observed FUV flux and the minimum measured $\rho_{HI}$ (Equation \ref{eqn:G0}). Taking the respective maximum and minimum values from our results yields a maximum \Gnaught\ of 140. The actual maximum \Gnaught\ in our dataset is 62. Obtaining a combination of a high \Gnaught\ and a low $n$ is limited by the maximum HI column density that is observed.

Next, we used the HI column densities to calculate the area filling factor of M81. It is worth mentioning here that Figure \ref{fig:Rnorm_HI} weakly reflects the same highs and lows that feature in \citet{bra1997} (their Figure 6a) when they used ring averaged values. The individual HI patches, presumably indicating GMCs, follow the global trend in HI column densities. However, the HI columns of individual GMCs show too much scatter to reproduce the two peaks in \citet{bra1997}'s results. From the individual HI column measurements and the global radial distribution of HI in M81, the filling factor in Figure \ref{fig:Rnorm_FF} was derived. Table \ref{tab:RR} lists the maximum measured HI columns per 1 kpc interval, the equivalent surface densities, the corresponding surface densities of M81's radial profile, and the resulting area filling factor. The radial profile has a tilted ring correction, appropriate for an inclined disk. At a local scale this correction is not appropriate, since the morphology of the individual HI columns is unknown. 
Note that we have no sources between 0.95 and 1.11 $\times\ R_{25}$ (resp. 12 and 14 kpc). The average filling factor within $R_{25}$ is 0.21. 

\subsection{Far Ultraviolet Flux}
Table \ref{tab:FUVfluxes} lists the locations and FUV fluxes of the central sources of our 27 selected regions. The candidate PDRs were selected to have a range of FUV fluxes and galactocentric radii. We did not find candidate PDRs in the inner region of M81 ($R/R_{25} < 0.3)$ with an FUV flux above $0.15 \times$ \UVunits. Qualitatively, neither detectable FUV emission nor detectable HI columns are present there. 

$F_{FUV}$ as shown in Figure \ref{fig:Rnorm_F} shows a weak correlation with galactocentric radius. No values of the FUV flux of individual sources larger than $11.56 \times$ \UVunits\ were found. At larger galactocentric radius, the UV sources become less bright. This could be caused by fewer dissociating stars per cluster or a change in the stellar population (an increasing amount of B stars and a decreasing amount of O stars). At the lower end of the FUV flux values, the minimum measured FUV flux values (and \Gnaught, Figure \ref{fig:Rnorm_G0}) seem to decrease with increasing galactocentric radius. This is potentially a confusion effect, if the brighter sources raise the ambient UV flux, outshining the weaker sources. However, it could also be the effect that is responsible for the decreasing maximum UV fluxes. At this point we are unable to make that distinction.

Table \ref{tab:NHI} shows the ratio of the FUV flux to the FUV background level ($G/G_{bg}$). A value of 1 means that the FUV source appears to be equally bright as the FUV background as ``seen'' by the HI patch. Values range from a source flux of 1\% of the background level to over 23 times the background level.

Figure \ref{fig:Rnorm_G0} shows the incident flux on all of our candidate PDRs. The values of \Gnaught\ show a tendency to decrease with increasing galactocentric radius. \Gnaught\ varies over a range of 4 dex, while the FUV flux only varies over 1 dex and shows only a very weak trend (Figure \ref{fig:Rnorm_F}). The same holds for the values of $\rho_{HI}$ as plotted in Figure \ref{fig:Rnorm_rho}, ranging from 35 pc to almost 600 pc. Taken together, they are responsible for the trend in the values of \Gnaught.

\subsection{Total Hydrogen Volume Density}

Figure \ref{fig:Rnorm_n} shows the resulting total hydrogen volume density using the \citet{sta1984} dust-to-gas ratios (see Equation \ref{eqn:dd0}). Values range between 1 and 200 per cm$^3$, without a clear gradient.
Using the \citet{gar1987} dust-to-gas model (not plotted), the resulting volume densities are in the same range, but with larger values in the outer regions of M81 and smaller values in the inner parts. Even then a clear gradient is absent. On average the relative error of the resulting $n$ in our method is 88\%; see Appendix \ref{app:error} for a more complete error analysis.

Figure \ref{fig:selection} illustrates the various observational selection effects by showing the range of \Gnaught\ and $n$ that are observable with our method. The different limits are indicated by the Roman numerals I through V. These limits arise as follows:

\makeatletter
\renewcommand{\theenumi}{\Roman{enumi}}
\renewcommand{\labelenumi}{\theenumi.}
\makeatother
\begin{enumerate}
  \item HI column density (sensitivity) lower limit of $1 \times 10^{20}\ \mbox{cm}^{-2}$ at a characteristic \dtg\ of 1. 
  \item HI upper limit of $5 \times 10^{21}\ \mbox{cm}^{-2}$ at a characteristic \dtg\ of 0.9, due to the HI column becoming optically thick.
  \item The lowest total hydrogen volume density that we can observe. It is determined by the lowest observable column density divided by the diameter of the radio beam and is roughly 0.5 cm$^{-3}$. This limit is therefore a combination of a sensitivity limit and a resolution limit.
  \item The minimum \Gnaught\ (0.01) we can reasonably use, as mentioned previously.
  \item The maximum \Gnaught\ of 140 that we can obtain, also mentioned earlier.
\end{enumerate}

At the top end we have a theoretical limit of roughly $1 \times 10^5\ \mbox{cm}^{-3}$, a combination of the highest \Gnaught\ and the lowest HI column density (not shown). These values will not be observed due to points I and V above. Moreover, recall that Equation \ref{eqn:n} is valid only for values of $n < 10^4\ \mbox{cm}^{-3}$.
More common values of \Gnaught\ combined with low HI columns yield values of up to 1000 cm$^{-3}$. In short, the range of values in our results is bounded by observational selection, and the apparent trend in Figure \ref{fig:selection} must be considered a result of that selection.

\subsection{Polycyclic Aromatic Hydrocarbons}

We now turn to the additional evidence supporting the idea that the regions we selected are indeed PDRs. Morphologically, all sources are similar in that HI and PAHs can be found in the neighborhood of the UV emission. This is apparent from the detailed plots of the candidate PDRs in Figure \ref{fig:panels}. This apparent spatial connection at high resolution between the atomic hydrogen and UV emission in M81 was previously reported in \citet{all1997}. Almost all of our sources show PAH emission nearby. At the largest galactocentric radii, the PAH emission becomes too faint to detect.  Table \ref{tab:NHI} (last column) shows whether the measured HI patches have PAH emission peaks within 6\arcsec, or 100 pc. 31 out of 56 HI patches have PAH emission peaks within that range.

Our subtracted PAH map is contaminated by non-stellar non-PAH emission (hot dust). In regions of high radiation intensity, one will observe non-stellar emission from HII regions in the 3.6 $\mu$m image, but at a separation of a few arcseconds the general view is that the non-stellar emission is coming from PAHs (Willner, private communication). Despite the contaminations in the resulting image, we are only interested in a qualitative comparison, where the presence of emission is more important than the actual amounts that occur.  PAHs are thought to be a tracer of recent star formation \citep{pee2004}. In that case, the presence of PAH emission is a further confirmation that the physical conditions typical of PDRs are indeed present.

With the exception of 4 sources at higher galactocentric radius that did not show detectable PAH emission (in general, PAH emission levels drop with increasing galactocentric radius), all UV sources have accompanying PAH emission.
Owing to differences in excitation conditions, we did not expect PAH emission to appear at the exact location of HI emission. The majority of HI patches, however, has PAH emission associated to it, suggesting a close relation between the two. A closer investigation of this correlation appears desirable, but is beyond the scope of this paper.

\begin{deluxetable}{rcccc}
\tabletypesize{\scriptsize}
\tablecaption{HI surface density per radius interval and resulting filling factor.}
\tablewidth{0pt}
\tablehead{
\colhead{Radius} & \colhead{Max. HI column} & \colhead{Equivalent $\Sigma_{HI}$} & \colhead{Radial $\Sigma_{HI}$} & \colhead{Filling Factor} \\
\colhead{(kpc)} & \colhead{\HIunits} & \colhead{($M_\odot$ pc$^{-2}$)} & \colhead{($M_\odot$ pc$^{-2}$)} & \colhead{}
}
\startdata
4-5~  & 2.91 & 22.17 & 3.88 & 0.17 \\
5-6~  & 4.83 & 36.79 & 4.80 & 0.13 \\
6-7~  & 2.88 & 21.94 & 4.62 & 0.21 \\
7-8~  & 3.45 & 26.28 & 6.41 & 0.24 \\
8-9~  & 4.58 & 34.89 & 5.56 & 0.16 \\
9-10  & 2.74 & 20.87 & 5.31 & 0.25 \\
10-11 & 3.20 & 24.37 & 5.42 & 0.22 \\
11-12 & 2.44 & 18.59 & 4.97 & 0.27 \\
14-15 & 2.83 & 21.56 & 2.67 & 0.12 \\
\enddata
\label{tab:RR}
\end{deluxetable}

\section{Discussion and Conclusions}
\label{sec:discussionconclusions}

The candidate PDRs in M81, which were selected on the basis of their FUV emission, seem to fit the photodissociation model well. Our results show no systematically different properties of the parent GMCs in different parts of M81. The total hydrogen volume density is roughly constant, even as the underlying HI, FUV and dust-to-gas ratio vary.

The cloud densities we find are lower than the range of values (30 - 1000 cm$^{-3}$) found in M101 \citep{smi2000}. The observed values of $N_{HI}$ of individual PDRs are similar to those seen in M101. The observed HI columns in both galaxies abruptly increase at the same normalized radius (0.3) and appear to decline somewhat beyond 0.7, assuming an $R_{25}$ of 30.3 kpc for M101 \citep{vil1992}.  The range of observed HI columns is also the same. The downward trend in \Gnaught\ also is consistent with \citet{smi2000}, when no internal extinction correction is applied to their data.
Figure \ref{fig:Rnorm_n} shows that the densities of the GMCs in M81 do not appear to change with galactocentric radius, consistent with the M101 results.

\citet{kau1999} modeled the expected CO intensity from a PDR for a range of incident FUV fluxes and cloud densities. The range of our results would be consistent with modeled CO intensities below $5 \times 10^{-8}~ \mbox{ergs cm}^{-2}~ \mbox{s}^{-1}~ \mbox{sr}^{-1}$, or $1 \times 10^{-8}~ \mbox{ergs cm}^{-2}~ \mbox{s}^{-1}~ \mbox{sr}^{-1}$ for the vast majority of sources (6.4 K km s$^{-1}$). Figure 1 in \citet{all2004} shows that for the range of values in Figure \ref{fig:selection}, the modeled CO intensities are independent of $n$. The low volume densities we find are consistent with a lack of CO emission in M81 as discussed by \citet{kna2006}, and which those authors attribute to insufficient excitation. We note that the volume number densities of colliding \Htwo\ molecules in the GMCs of M81 is 1/2 of the values for $n$ calculated here. The mean value we have found for $n$ in the GMCs of M81 translates into a mean number density for $n_{H_2} \approx 10 \mbox{cm}^{-3}$, well below the values required for collisional excitation of CO molecules. In this case, the CO emission in M81 is in general \textit{subthermal}. Their reported upper limit for $I_{CO}$ is 1.03 K km s$^{-1}$ for the regions they investigated, near our source no. 18. The \Gnaught\ and $n$ that we find there are consistent to that value of $I_{CO}$ (and somewhat higher) per Figure 1 in \citet{all2004}, when beam dilution is taken into account.
The low CO emission in M81 is also explored in \citet{cas2007}, who again point to a lack of excitation of the molecular gas. They find no molecular gas in the nucleus of M81. The absence of FUV sources to excite the gas is consistent with that finding.

After accounting for observational and projection effects (see the previous Section), we note that in the nearby galaxies in general on which our method is applicable, it is most sensitive to a combination of low \Gnaught\ and low $n$.

Summarizing, our conclusions are:

\begin{itemize}
\item We selected a number of discrete FUV sources in M81, which we consider to be potential PDRs on the surfaces of the parent GMCs.
\item The total hydrogen volume densities of GMCs close to clusters of young stars in M81 are in the range of 1 $< n <$ 200 cm$^{-3}$ with a geometric mean of 17 cm$^{-3}$. This is approximately ten times lower than GMCs in M101 studied with the same method.
\item The low GMC volume densities are consistent with a lack of CO emission in M81. 
\item M81's GMCs have a filling factor of $\approx$ 0.21 within $R_{25}$.
\item No candidate PDRs are found in M81 within $R < 0.3 R_{25}$.
\item We have provided a thorough analysis of the observational selection effects on our results and conclude that, while such effects are (necessarily) present in our results, our main conclusions as to the range and values of the total volume densities of GMCs in M81 are not affected.
\item The presence of PAH emission in the neighborhood of our candidate PDRs lends support to our view that the HI patches near FUV sources are indeed produced by photodissociation. PAH emission occurs near almost all UV sources. PAH and HI emission coincide in more than half of our sources.
\end{itemize}

\acknowledgements
The authors thank R. Braun for providing the VLA data we used initially, yielding valuable insights into data reduction and the noise levels of M81's HI data. Thanks go to E. Brinks and F. Walter, for making available their excellent THINGS M81 data and for invaluable discussions as well as helpful comments regarding earlier drafts of this paper. We are grateful to Frank Bigiel for providing us with the THINGS M81 radial profile. We thank G. Meurer for helping with the GALEX data. We also thank J. Morrison, K. Sahu, B. McLean and R. Bohlin for helping out with astrometrical corrections and analysis of the UV data. We thank N. Panagia and X. Tielens for advice and useful discussions about the physics of PDRs. M. Vogelaar has been extremely helpful with resolving some specific problems peculiar to the data set we could obtain. We thank B. Holwerda for helping out with preliminary versions of the M81 Spitzer data from the archives. The anonymous referee provided many useful suggestions, for which we are grateful. Finally we thank J. Knapen for useful general discussions about M81.

JSH acknowledges the support of a Graduate Research Assistantship provided by the STScI Director's Discretionary Research Fund.

\appendix

\section{Error analysis}
\label{app:error}

To estimate the extent to which the different observables influence the result, we performed a basic error analysis.  It takes into account the precision of the determination of the UV flux $F$, the HI-to-UV distance $\rho$, the dust-to gas ratio $\delta/\delta_0$ and the HI column densities ($N$). Note that in this case we subsitute \Gnaught\ with its observables $F$ and $\rho$.

Formula \ref{eqn:n} for $n$ can be written as 

\begin{equation}
n = C \frac{F}{\rho^2} \left(\delta/\delta_0\right)^{-1/2}\left[\exp{(N\delta/\delta_0)} - 1\right]^{-1},
\label{eqn:ncst}
\end{equation}
where $C$ is a constant that includes the extinction, the distance to M81, the scaling factor for the flux and the other constant factors that went into the model. Also note that $N \equiv N_{HI}/(7.8 \times 10^{20}\ \rm{cm^{-2}})$.

\begin{eqnarray}
\sigma_n^2 &=& 
\left(\frac{\partial n}{\partial F} \right)^2 \sigma_F^2 +
\left(\frac{\partial n}{\partial \rho} \right)^2 \sigma_\rho^2 + \nonumber \\
&& \left(\frac{\partial n}{\partial \delta} \right)^2 \sigma_\delta^2 +
\left(\frac{\partial n}{\partial N} \right)^2 \sigma_N^2,
\label{eqn:nerr}
\end{eqnarray}
the standard formula for propagation of uncertainties, where

\begin{eqnarray}
\frac{\partial n}{\partial F} &=& \frac{n}{F} \\
\frac{\partial n}{\partial \rho} &=& -2 \frac{n}{\rho} \\
\frac{\partial n}{\partial \delta} &=& -\frac{1}{2} \left(\delta/\delta_0\right)^{-1} n - \nonumber \\ && n \left[\exp{(N \delta/\delta_0)} - 1 \right]^{-1} N \exp{(N \delta/\delta_0)} \\
\frac{\partial n}{\partial N} &=& -n \left[\exp{(N \delta/\delta_0)} - 1\right]^{-1} \cdot \nonumber \\
&& \delta/\delta_0 \exp{(N \delta/\delta_0)}
\end{eqnarray}

The square root of the result, $\sigma_n$, is the standard deviation, or the error. The dust-to-gas ratio has a disproportionate influence on the error, since it occurs twice in the equation and is also the parameter that is the least well known.
The relative precision of the measurements of $F$ and $N$ is roughly 1\%. The separation has an uncertainty in the order of a half a pixel, or about 9 pc. Finally, the scatter in the fit to the dust-to-gas ratio is about 0.2.

Taken together, these figures introduce relative errors to the result of about 88\% of the measured value of $n$ on average, or approximately a factor of 2. This means that for example $n = 1 \pm 0.8$ if $\frac{\sigma_n}{n} = 80\%$. Negative values of $n$ are unphysical.
The spread in our resulting values of $n$ seems to reflect this and we expect the accuracy of our results to be similar.


\end{document}